\documentclass[conference,letterpaper,final]{IEEEtran}
\usepackage{geometry}
\usepackage{color}
\usepackage{amsmath}
\usepackage{amsthm}
\usepackage{amssymb}
\usepackage{graphicx}
\usepackage{mathtools}

\usepackage{setspace}
\usepackage{url}

\usepackage{cite}
\usepackage{epsfig}
\usepackage{anysize}
\usepackage{algorithm}
\usepackage{algorithmic}
\usepackage{comment}
\usepackage{setspace} %[nodisplayskipstretch]
\makeatletter

\theoremstyle{plain}

\theoremstyle{plain}
\newtheorem{prop}{\protect\propositionname}
\theoremstyle{plain}

\theoremstyle{plain}

\theoremstyle{remark}

\marginsize{0.6in}{0.6in}{0.35in}{.35in}

\makeatother

\providecommand{\remarkname}{Remark}
\providecommand{\lemmaname}{Lemma}
\providecommand{\corollaryname}{Corollary}
\providecommand{\propositionname}{Proposition}
\providecommand{\theoremname}{Theorem}

\allowdisplaybreaks
\begin{document}

%\title{Performance Analysis of Cache-Aided\\ Massive MIMO}
%\title{Cache-Aided Massive MIMO for Multi-user Downlink Transmission}
\title{Cache-Aided Massive MIMO: Linear Precoding Design and Performance Analysis}
%\author{Xiao Wei$^{*\ddagger}$, Lin Xiang$^{\S}$, Laura Cottatellucci$^{\ddagger}$, Robert Schober$^{\ddagger}$, and Tao Jiang$^*$\\
%$^*$School of Electronics Information and Communications, Huazhong University of Science and Technology, China\\
%$^{\ddagger}$Institute for Digital Communications, Friedrich-Alexander-University Erlangen-N\"urnberg, Germany\\
%$^{\S}$Interdisciplinary Centre for Security, Reliability and Trust (SnT), University of Luxembourg, Luxembourg\\
%Email: weixiao1991@hust.edu.cn; lin.xiang@uni.lu; \{laura.cottatellucci, robert.schober\}@fau.de; Tao.Jiang@ieee.org}

\author{Xiao Wei$^{*\ddagger}$, Lin Xiang$^{\S}$, Laura Cottatellucci$^{\ddagger}$, Tao Jiang$^*$, and Robert Schober$^{\ddagger}$\\
$^*$Huazhong University of Science and Technology, P.R. China;
$^{\S}$University of Luxembourg, Luxembourg;\\
$^{\ddagger}$Friedrich-Alexander-University Erlangen-N\"urnberg, Germany.
}

\maketitle
\begin{abstract}
In this paper, we propose a novel joint caching and massive multiple-input multiple-output (MIMO) transmission scheme, referred to as cache-aided massive MIMO, for advanced downlink cellular communications. In addition to reaping the conventional advantages of caching and massive MIMO, the proposed scheme also exploits the side information provided by cached files for interference cancellation at the receivers. This interference cancellation increases the degrees of freedom available for precoding design.
In addition, the power freed by the cache-enabled offloading can benefit the transmissions to the users requesting non-cached files. The resulting performance gains are not possible if caching and massive MIMO are designed separately. We analyze the performance of cache-aided massive MIMO for cache-dependent maximum-ratio transmission (MRT), zero-forcing (ZF) precoding, and regularized zero-forcing (RZF) precoding. Lower bounds on the ergodic achievable rates are derived in closed form for MRT and ZF precoding. The ergodic achievable rate of RZF precoding is obtained for the case when the numbers of transmit antennas and users are large but their ratio is fixed.
Compared to \emph{conventional} massive MIMO, the proposed cache-aided massive MIMO scheme achieves a significantly higher ergodic rate especially when the number of users approaches the number of transmit antennas.
\end{abstract}

\section{Introduction}
Massive multiple-input multiple-output (MIMO) is a key technology to improve the spectral efficiency of cellular communications and thus, to support the explosive growth of cellular traffic \cite{Wong2017, N2013}.
By employing a large number of antennas at the base station (BS), massive MIMO offers abundant spatial degrees of freedom and facilitates large multiplexing and diversity gains.

For a single-cell massive MIMO system, where the BS is equipped with $M$ antennas and communicates with $K$ single-antenna users, performance has been shown to depend critically on the number of BS antennas per user, denoted as $\rho_0 \triangleq M/K$ \cite{Huh2011}. For example, when both $M$ and $K$ grow without bound while $\rho_0$ is finite, the authors of \cite{Yang2013} show that the effective signal-to-interference-plus-noise ratio $\!$ (SINR) grows linearly with $\rho_0$. In \cite{Ngo2013}, the authors show that simple linear precoders and detectors are asymptotically optimal, i.e., capacity-achieving, in the large system limit where $M$ grows unbounded while $K$ is fixed, i.e., $\rho_0$ becomes very large.
Since the publication of \cite{Yang2013} and \cite{Ngo2013}, massive MIMO has been typically studied for low-complexity linear precoding \cite{Raeesi2018}, which performs well for large $\rho_0$.
However, due to emerging applications, related to e.g. smart phones, tablets, and Internet-of-Things devices, in future wireless systems, the number of users may grow significantly, even beyond the number of BS antennas \cite{Yu2017}.
In this case, conventional linear precoding based massive MIMO suffers from a significant performance loss due to the resulting small $\rho_0$. Hence, improving the performance of massive MIMO systems when $\rho_0$ is small is important for future applications but has not been sufficiently addressed in the literature \cite{Bjornson2018}.

In this paper, we show that wireless caching at the user side provides an opportunity for enhancing the capacity of massive MIMO, especially for small $\rho_0$. With the wide-spread use of smart phones and tablets, cache memory is often available at the users.
By pre-storing the most popular files in the users' caches during periods of low network traffic, fast access to these files is enabled without requiring over-the-air delivery \cite{Semiari2018}.
However, as the actual users' requests are not known during cache placement, the cached files may not be requested by the users later on, which severely limits the performance gains of user-side caching.
To mitigate this problem, two approaches, which exploit additional performance gains enabled by caching, have been proposed in the literature \cite{Xiang2018, Maddah-Ali2014}.
One approach, referred to as cache-aided non-orthogonal multiple access (NOMA) \cite{Xiang2018}, exploits a user's cached but non-requested files for canceling NOMA interference. By joint optimization of cache-enabled interference cancellation and successive interference cancellation, cache-aided NOMA can significantly improve the users' achievable rates \cite{Xiang2018}. However, when the number of users is large, the optimization of cache-aided NOMA becomes intractable. An alternative approach employs coded caching \cite{Maddah-Ali2014}. By carefully encoding the cached and delivered files, simultaneous multicast to multiple users is enabled such that each user can decode its requested file without suffering from multiuser interference \cite{Maddah-Ali2014}.
However, for coded caching, forming multicast groups and decoding impose a large computational burden on both the BS and the users \cite{Maddah-Ali2014}. Synergies between caching and massive MIMO are explored in \cite{Ngo2018}, where several communication schemes are proposed and analyzed. One of the schemes in \cite{Ngo2018} combines coded caching with massive MIMO for improved multicast transmission over fading channels, whereas another scheme leverages the spatial multiplexing capability of massive MIMO to simultaneously transmit the requested files to all the users. Finally, a combination of the two schemes is also analyzed in \cite{Ngo2018}.

In this paper, we propose several novel cache-aided precoding schemes for massive MIMO, which are collectively referred to as cache-aided massive MIMO. Assume that each user is equipped with a cache memory and the BS is equipped with a large number of antennas. If the files cached at one user are requested by the user itself, caching offloads the transmission to the user and, by cache-aided massive MIMO, more transmit power can be allocated to the other users. On the other hand, if the files cached at a user are not requested by the user itself but are requested by other users, these files can be exploited for interference cancellation at the user, avoiding the need for interference suppression via precoding at the BS. Consequently, cache-aided massive MIMO introduces additional degrees of freedom for the transmission of the remaining files which leads to improved performance.
Appealingly, owing to the large antenna array at the BS, the performance gains enabled by caching are achievable via properly redesigned linear precoders.
Hence, cache-aided massive MIMO avoids the encoding and decoding overhead incurred by coded caching [11] and cache-aided NOMA [9], and is computationally efficient.
The main contributions of this paper are as follows:
\begin{itemize}
\item We propose a novel cache-aided massive MIMO scheme, which not only facilitates offloading and interference cancellation at the users, but also enhances the precoding at the BS.

\item To reap these cache-enabled benefits, low-complexity linear precoders based on maximum-ratio transmission (MRT), zero-forcing (ZF), and regularized zero-forcing (RZF) precoding are proposed. Different from conventional linear precoding which only requires channel state information (CSI) at the BS, the proposed linear precoders depend on both the CSI and the cache status. We analyze the performance of cache-aided massive MIMO for each considered linear precoding scheme. Lower bounds on the ergodic achievable rates of MRT and ZF precoding are derived in closed form. Additionally, we analyze the asymptotic performance of RZF precoding based on random matrix theory.

\item Simulation results show that, compared to \emph{conventional} massive MIMO, the proposed cache-aided massive MIMO scheme achieves a significantly higher ergodic rate, especially when $\rho_0$ is small.
\end{itemize}

We note that for notational convenience, we only consider a simple caching policy where each user caches entire files. However, the proposed precoding techniques are applicable to more general caching policies and may achieve even higher throughputs if each user caches portions of each file as in \cite{Maddah-Ali2014, Ngo2018}.
Due to the limited page space, the design of throughput-optimal caching policies for the proposed cache-aided massive MIMO scheme and the corresponding performance analysis are deferred to future work.

The remainder of the paper is organized as follows. In Section II, the system model and the proposed cache-aided massive MIMO scheme are presented. We analyze the achievable rates of the proposed scheme for different linear precoders in Section III. The performance of cache-aided massive MIMO is evaluated in Section IV, and finally, Section V concludes the paper.

\emph{Notations}: In this paper, we use boldface capital and lower case letters to denote matrices and vectors, respectively.
${\bf{A}}^{\rm H}$ and ${\bf{A}}^{\rm T}$ represent the complex conjugate transpose and the transpose of matrix $\bf{A}$, respectively;
${\bf{A}}^{-1}$ is the inverse of square matrix $\bf{A}$;
$\Pr(\cdot)$ and $\mathcal{E}\{\cdot\}$ are the probability and the expectation operators, respectively;
$\Re(\cdot)$ and $\Im(\cdot)$ represent the real and imaginary parts of a complex number, respectively.
$\parallel \cdot \parallel $ and $\mid \cdot \mid $ are the Euclidean norm of a vector and the absolute value of a scalar, respectively.
$C_n^k$ is the $k$-out-of-$n$ binomial coefficient.
$A \rightarrow B$ indicates that $A$ converges to $B$ in the limit.
$\mathcal{CN} \left(0,\sigma^2\right)$ denotes the complex Gaussian distribution with zero mean and variance $\sigma^2$,
and finally, ${{\mathbb C}^{N_r \times N_t}}$ is the set of complex-valued $N_r \times N_t$ matrices.

\section{Cache-Aided Massive MIMO}

In this section, the system model, the interference cancellation mechanism, and the ergodic achievable rate of the proposed cache-aided massive MIMO scheme are presented.

\subsection{System Model}
\begin{figure}
\centering
\includegraphics[scale=0.35]{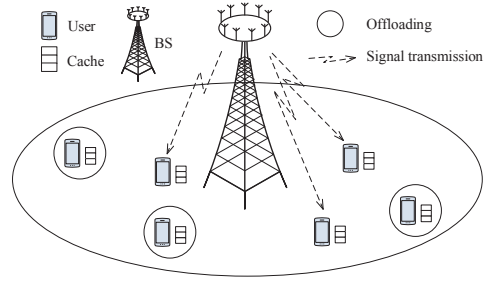}
\caption{Delivery model for cache-aided massive MIMO system, where each user obtains the requested file from its cache or the BS, depending on its cache status.}
\vspace{-0.5em}
\label{fig0.1}
\end{figure}

As shown in Fig.~\ref{fig0.1}, we consider a single-cell downlink system with an $M$-antenna BS and $K$ single-antenna users. The BS stores a library of $L_b$ popular files, where each file has a size of $F$ MBytes. Each user is equipped with a cache memory of size $L_uF$ MBytes, where $L_u \leq L_b$, i.e., the cache capacity of each user is insufficient to store the whole library and only a portion of the files can be cached.

The system operates in two phases: a placement phase and a delivery phase. In the placement phase, all the users place $L_u$ arbitrary files from the library into their own cache prior to the time of request. This phase may happen during the early mornings when cellular traffic is low. In the delivery phase, each user may request one of the $L_b$ files. Let $c_{k,l}=0$ if the file requested by user $k$ has been cached at user $l$ and $c_{k,l}=1$ otherwise. Using this notation, the number of active transmission users in the proposed system is given by $\overline{K}=\sum\nolimits_{k=1}^K c_{k,k}$. If user $k$ has cached the file it requests, i.e., $c_{k,k}=0$, it is fetched from its cache instantly. In this case, user $k$ is considered inactive as it requires no cellular transmission. Otherwise, if $c_{k,k}=1$, the requested file has to be transmitted by the BS, and user $k$ is considered to be active.

%Moreover, transmission design can further enhance these performance by combining the transmission design at the BS, such as precoding.

\subsection{Cache-Enabled Interference Cancellation}
In this paper, we assume that the cache status is given and we focus on exploiting the cached data to improve the delivery to all the users. Assume that the file requested by user $k$ is not cached at user $k$, i.e., $\!c_{k,k}\!\!=\!\!1\!$ and the BS has to transmit the file. Then, the received signal at user $k$, denoted by $y_{k}$, is given by
\begin{align}\label{eq2.0}
y_{k}={\bf{h}}_{k}^{\rm H}{\bf{w}}_{k}s_{k}+\sum\limits_{l\neq k} c_{l,l} {\bf{h}}_{k}^{\rm H}{\bf{w}}_{l}s_{l}+v_{k},
\end{align}
where $s_{k}$ is the transmit symbol intended for user $k$ with $\mathcal{E} \{ \left| s_k\right| ^2 \}=E_k$. ${\bf{w}}_k\!\in\!{{\mathbb C}^{{M} \times {1}}}$ is the precoding vector of user $k$ and $v_k$ is the additive white Gaussian noise following distribution $\mathcal{CN} \left(0,\sigma^2\right)$.
${\bf{h}}_{k}=\left[h_{k,1},h_{k,2},\cdots,h_{k,M}\right]^{\rm T}\!\in\!{{\mathbb C}^{{M} \times {1}}}$ is the channel vector from the BS to  user $k$.
In this paper, channel coefficient $h_{k,m}$ is modeled as
\begin{equation}\label{eq1.2}
h_{k,m} = g_{k,m}\sqrt{\beta_{k}},
\end{equation}
where $g_{k,m}$ is the fading coefficient from the $m$th BS antenna to user $k$ and follows distribution $\mathcal{CN}(0,1)$.
$\beta_{k}$ models the pathloss and shadowing effects and remains constant over a large number of coherence time intervals. We assume that the total transmit power at the BS is $E_0$. To satisfy the total transmit power constraint, we let $ \sum\nolimits_{k=1}^K \!c_{k,k}E_k \! = \! E_0$ and $\left\| {\bf{w}}_k \right\| ^2\!=\!1$, $\forall k$.

If $c_{l,l}=0$, user $l$ is inactive and hence, it is offloaded for transmission in \eqref{eq2.0}. On the other hand, if $c_{l,l}=1$ and $c_{l,k}=0$, i.e., the file requested by active user $l$ is not cached at user $l$ but is cached at user $k$, this cached file can still be exploited for interference cancellation \cite{Xiang2018}.
In particular, by re-encoding this cached file and subtracting the corresponding signal from $y_k$, the interference caused by user $l$ to user $k$ can be removed\footnote{For suppressing the interference caused by user $l$ at user $k$, the re-encoded signal $s_l$ is scaled by ${\bf{h}}_{k}^{\rm H}{\bf{w}}_{l}$ before being subtracted from the received signal. Here, ${\bf{h}}_{k}^{\rm H}{\bf{w}}_{l}$ can be estimated locally at user $k$ requiring no knowledge about the requests and cache status of the other users.} at user $k$.
Consequently, by caching at the user side, user $l$ with $c_{l,l} = 1$ causes interference only to users $k$ with $c_{l,k} =1$, $l\neq k$.
Let $U_k \triangleq \left\{ l \mid c_{l,l} = 1, c_{l,k} = 1, l\neq k  \right\}$ be the set of users interfering with user $k$. The cardinality of $U_k$ is denoted by $N_k$. The residual received signal of user $k$ after interference cancellation, denoted by $y_{k}^{\mathrm{IC}}$, is given by
\begin{align}\label{eq2.1}
y_{k}^{\mathrm{IC}}={\bf{h}}_{k}^{\rm H}{\bf{w}}_{k}s_{k}+\sum\limits_{l\in U_k} {\bf{h}}_{k}^{\rm H}{\bf{w}}_{l}s_{l}+v_{k}.
\end{align}

\subsection{Ergodic Achievable Rate}
Based on \eqref{eq2.1}, the SINR of user $k$ is given by
%\begin{align}\label{eq2.3}
%\mathrm{SINR}_{k}&=\frac{\mid {\bf{h}}_{k}^{\rm H}{\bf{w}}_{k} s_{k}\mid^2}{\sum\limits_{l\in U_k}\mid {\bf{h}}_{k}^{\rm H}{\bf{w}}_{l} s_l\mid^2+\mid v_{k} \mid^2} \nonumber \\
%&=\frac{\mid {\bf{h}}_{k}^{\rm H}{\bf{w}}_{k}\mid^2E_{k}}{\sum\limits_{l\in U_k}\mid  {\bf{h}}_{k}^{\rm H}{\bf{w}}_{l}\mid^2E_{l}+\sigma^2}.
%\end{align}
\begin{align}\label{eq2.3}
\mathrm{SINR}_{k}=\frac{\left| {\bf{h}}_{k}^{\rm H}{\bf{w}}_{k}\right|^2E_{k}}{\sum\limits_{l\in U_k}\left|  {\bf{h}}_{k}^{\rm H}{\bf{w}}_{l}\right|^2E_{l}+\sigma^2}.
\end{align}
Assume that the file delivery for each active user spans a large number of coherence time intervals. Then, with the proposed cache-aided massive MIMO, the ergodic achievable rate of user $k$ is \cite{N2013}
\begin{align}\label{eq2.4}
R_{k}&=\mathcal{E}\left\{\log_2\left(1+\mathrm{SINR}_{k}\right)\right\} \nonumber \\
&=\mathcal{E}\left\{\log_2\left(1+\frac{\left| {\bf{h}}_{k}^{\rm H}{\bf{w}}_{k}\right|^2 E_k}{\sum\limits_{l\in U_k}\left| {\bf{h}}_{k}^{\rm H}{\bf{w}}_{l}\right|^2 E_l+\sigma^2}\right)\right\}.
\end{align}
Since $f(x)=\log_2(1+\frac{1}{x})$ is a convex function, by employing Jensen's inequality, a lower bound on the ergodic achievable rate $R_k$ is obtained as \cite{N2013}
\begin{equation}\label{eq2.5}
R_{k}\!\geq\! \tilde{R}_{k}\!\stackrel{\Delta}{=}\!\log_2\!\left(\!1\!+\!\left(\mathcal{E}\left\{\frac{\sum\limits_{l\in U_k}\left| {\bf{h}}_{k}^{\rm H}{\bf{w}}_{l}\right|^2 E_l\!+\!\sigma^2}{\left| {\bf{h}}_{k}^{\rm H}{\bf{w}}_{k}\right|^2 E_k}\right\}\right)^{\!\!-1}\!\right)\!.\!
\end{equation}

\section{Cache-Aided Linear Precoder Design}
In this section, we investigate advanced precoder designs at the BS exploiting user-side caching for enhanced performance. We consider linear precoding techniques, namely MRT, ZF, and RZF, which are preferred in practical massive MIMO systems as they attain high performance with affordable computational complexity. We analyze the ergodic achievable rates for MRT and ZF precoding and derive corresponding lower bounds in Sections III-A and III-B, respectively. In Section III-C, we analyze the achievable rate for RZF precoding in the asymptotic regime, where $M, K \rightarrow \infty$ but $\rho_0$ is fixed. We assume that the CSI, ${\bf{h}}_k$, is perfectly known at both the receivers and the transmitter. Here, we consider $M>K$, while the proposed scheme is also applicable for $M\leq K$.

\subsection{Maximum-Ratio Transmission}
MRT precoding ensures that the signals transmitted by the BS over different antennas add up constructively at the intended user, and hence, maximizes the received signal power.
However, when $\rho_0$ is small, MRT suffers from severe multiuser interference.
In this case, cache-aided interference cancellation and offloading can improve the performance of MRT.
With MRT and perfect CSI, the precoding vector of user $k$ is \cite{Yang2013}
\begin{equation}\label{eq2.6}
{\bf{w}}_k^{\mathrm{MRT}}=\frac{{\bf{h}}_k}{\left\| {\bf{h}}_k \right\|}.
\end{equation}
By substituting \eqref{eq2.6} into \eqref{eq2.4}, the ergodic achievable rate of user $k$ with MRT precoding is
\begin{equation}\label{eq2.6.2}
R_{k}^{\mathrm{MRT}}=\mathcal{E}\left\{\log_2\left(1+\frac{ \left\| {\bf{h}}_{k} \right\|^2 E_k}{\sum\limits_{l\in U_k}
\frac{ \left| {\bf{h}}_{k}^{\mathrm H}{\bf{h}}_{l} \right|^2}{ \left\| {\bf{h}}_l \right\|^2} E_l+\sigma^2}\right)\right\}.
\end{equation}

\begin{prop} \label{prop1}
\emph{With MRT precoding in \eqref{eq2.6} and perfect CSI, the ergodic achievable rate of user $k$ is lower bounded as
\begin{equation}\label{eq2.7}
R_{k}^{\mathrm{MRT}} \geq \tilde{R}_{k}^{\mathrm{MRT}}=\log_2\left(1+\frac{\beta_k(M-1)E_k}{{\sum\limits_{l\in U_k}\beta_k E_l+\sigma^2}}\right).
\end{equation}
}
\end{prop}
\begin{IEEEproof}
Please refer to Appendix~\ref{proof1}.
\end{IEEEproof}

In Proposition~\ref{prop1}, fading is averaged out in $\tilde{R}_{k}^{\mathrm{MRT}}$. Moreover, as the desired and the interfering signals experience the same downlink channel, the lower bound on the ergodic achievable rate of a user depends only on the pathloss and shadowing of its own channel. The sum ergodic achievable rate of all users can be improved by optimizing the power allocation given the pathloss and shadowing of the different users. However, to illustrate the performance gains enabled by caching, we simply  assume uniform transmit power allocation for all active users, where $E_k=E_0/\overline{K}$.
Consequently, the lower bound in \eqref{eq2.7} simplifies to
\begin{equation}\label{eq2.8}
\tilde{R}_{k}^{\mathrm{MRT,uni}}=\log_2\left(1+\frac{\beta_k(M-1)E_0}{{\beta_k N_k  E_0+\overline{K}\sigma^2}}\right).
\end{equation}
Based on \eqref{eq2.8}, the ergodic achievable rate of \emph{conventional} massive MIMO (i.e., $N_k=K-1$, $\overline{K}=K$) for uniform transmit power allocation is lower bounded by
\begin{equation}\label{eq2.6.3}
\tilde{R}_{k}^{\mathrm{b,MRT,uni}}=\log_2\left(1+\frac{\beta_k(M-1)E_0}{{\beta_k (K-1) E_0+K\sigma^2}}\right).
\end{equation}
\emph{Remark} 1.
In \eqref{eq2.8}, $N_k$ and $\overline{K}$ are proportional to $K$. This implies that both $\tilde{R}_{k}^{\mathrm{MRT,uni}}$ and $\tilde{R}_{k}^{\mathrm{b,MRT,uni}}$ increase monotonically with the number of BS antennas per user. Moreover, we have $\tilde{R}_{k}^{\mathrm{MRT,uni}} \geq \tilde{R}_{k}^{\mathrm{b,MRT,uni}}$,
i.e., \eqref{eq2.6.3} defines the worst-case performance of the proposed scheme.
Comparing \eqref{eq2.8} and \eqref{eq2.6.3}, the performance gains of the proposed scheme over \emph{conventional} massive MIMO include: i) an enhanced transmit power for the active users due to cache-enabled offloading, when ${\overline{K}} < K$, and ii) reduced interference due to cache-enabled interference cancellation and offloading, when $N_k < K -1$.

\subsection{Zero-Forcing Precoding}
Different from MRT, ZF precoding avoids multiuser interference by projecting the transmit signal of each user into the null space of all other users. When files are cached at the users' terminals, the interference cancellation capabilities offered by caching and ZF precoding can be combined for improved precoding design.
In particular, if $c_{l,l}=0$, i.e., user $l \neq k$ is inactive, its channel will not be considered for the ZF precoding design at user $k$;
%user $k$ would not cause interference to user $l$;
on the other hand, if user $l$ is active and has cached the file requested by user $k$, i.e., $c_{l,l}=1$ and $c_{k,l}=0$, user $l$ can exploit the cached file to remove the interference caused by user $k$, without having to rely on ZF precoding. Hence, the ZF percoder intended for user $k$, only needs to avoid causing interference to the set of active users that do not have user $k$'s requested file in their caches. This set of users is denoted by $\Lambda_k \triangleq \left\{ l \mid c_{l,l}\!=\!1, c_{k,l}\!=\!1, l\neq k  \right\}$.
Consequently, if $\Lambda_k$ is not empty, the precoding vector ${\bf{w}}_k^{\mathrm{ZF}}$ of user $k$ has to satisfy the following constraints:
\begin{align}\label{eq3.1}
& \left\| {\bf{w}}_k^{\mathrm{ZF}} \right\|^2=1, \ \text{and} \ {\bf{h}}_l^{\mathrm H} {\bf{w}}_k^{\mathrm{ZF}}=0,{l\in \Lambda_k},
\end{align}
such that the signal of user $k$ is sent in the null space of the signal space formed by the users in $\Lambda_k$. Let $D_k$ and $\Lambda_k (n)$ be the cardinality and the $n$th element of set $\Lambda_k$, respectively. Then, for user $k$ and set $\Lambda_k$, we define the effective channel matrix after cache-enabled interference cancellation as
\begin{equation}\label{eq3.4}
{\bf{Q}}_k=[{\bf{q}}_{1},{\bf{q}}_{2},\cdots,{\bf{q}}_{D_k+1}],
\end{equation}
where ${\bf{q}}_{1}={\bf{h}}_k$, and ${\bf{q}}_{n+1}={\bf{h}}_{\Lambda_k (n)}$, $n = 1,2,\cdots,D_k$. %, $l\in \Lambda_k$.
Consequently, for the proposed cache-aided massive MIMO, the ZF precoding vector of user $k$ is given by
\begin{equation}\label{eq3.5}
{\bf{w}}_k^{\mathrm{ZF}}=\frac{{\bf{Q}}_k({\bf{Q}}_k^{\rm H}{\bf{Q}}_k)^{-1}{\bf{e}}_1}{ \left\| {\bf{Q}}_k({\bf{Q}}_k^{\rm H}{\bf{Q}}_k)^{-1}{\bf{e}}_1 \right\|},
\end{equation}
where ${\bf{e}}_1 \triangleq \left[1,0,\cdots,0 \right]^{\mathrm{T}}$. For the ZF precoder ${\bf{w}}_k^{\mathrm{ZF}}$, the corresponding ergodic achievable rate and its lower bound are given in Proposition~\ref{prop2}.

\begin{prop} \label{prop2}
\emph{
With the ZF precoder ${\bf{w}}_k^{\mathrm{ZF}}$ in (\ref{eq3.5}) and perfect CSI, the ergodic achievable rate of user $k$ is
\begin{equation}\label{eq3.3}
R_{k}^{\mathrm{ZF}}=\mathcal{E}\left\{\log_2\left(1+\frac{E_k }{{ \left\| {\mathbf{Q}}_k({\mathbf{Q}}_k^{\mathrm{H}}{\mathbf{Q}}_k)^{-1}{\mathbf{e}}_1 \right\|^2}\sigma^2}\right)\right\}.
\end{equation}
Moreover, $R_{k}^{\mathrm{ZF}}$ is lower bounded by
\begin{equation}\label{eq3.8}
R_{k}^{\mathrm{ZF}} \ge \tilde{R}_{k}^{\mathrm{ZF}}
=\log_2\left(1+\frac{\beta_k\left(M-D_k-1\right)E_k}{\sigma^2}\right).
\end{equation}
}
\end{prop}
\begin{IEEEproof}
Please refer to Appendix~\ref{proof2}.
\end{IEEEproof}

If uniform transmit power allocation is adopted, the lower bound in \eqref{eq3.8} simplifies to
\begin{equation}\label{eq3.9}
\tilde{R}_{k}^{\mathrm{ZF,uni}}=\log_2\left(1+\frac{\beta_k(M-D_k-1)E_0}{\overline{K}\sigma^2}\right).
\end{equation}
Based on (\ref{eq3.9}), the ergodic achievable rate of  user $k$ for \emph{conventional} massive MIMO (i.e., $D_k=K-1$, $\overline{K}=K$) is lower bounded by
\begin{equation}\label{eq3.3.1}
\tilde{R}_{k}^{\mathrm{b,ZF,uni}}=\log_2\left(1+\frac{\beta_k(M-K)E_0}{K\sigma^2}\right).
\end{equation}
\emph{Remark} 2.
We have $\tilde{R}_{k}^{\mathrm{ZF,uni}} \ge \tilde{R}_{k}^{\mathrm{b,ZF,uni}}$. To explain the performance difference, we note that, in the conventional ZF-based massive MIMO system, the signal of user $k$ has to be orthogonal to the signals of all other $K - 1$ users. The resulting interference mitigation comes at the cost of a reduced received signal power for each user. In contrast, with the proposed scheme, as cached-enabled offloading and interference cancellation can partially mitigate the interference, more spatial degrees of freedom are available for ZF precoding design and hence, the power loss incurred by ZF precoding is reduced. Moreover, due to cache-enabled offloading, a power gain of $K/{\overline{K}}$ is also achieved for transmit power allocation to the active users.

\subsection{Regularized Zero-Forcing Precoding}
In \emph{conventional} massive MIMO, RZF precoding is often considered to balance between interference mitigation and power enhancement \cite{Zhu2016}. For the proposed scheme, RZF precoding has to be reconsidered in order to maximize the performance gains enabled by caching. However, the ergodic rate of RZF precoding cannot be analyzed in the same manner as that of MRT/ZF precoding. To make the analysis tractable, we investigate RZF precoding in the large system limit, when $M, K \rightarrow \infty$ but $\rho_0$ is fixed and we assume that the $\beta_k$s are equal\footnote{This assumption facilitates concise and insightful results. Nevertheless, the extension to non-identical $\beta_k$s is possible \cite{Zhu2016} and will be provided in the journal version of the paper.}.

We define the effective channel fading matrix for user $k$ as
\begin{equation}\label{eq7.2}
{\bf{F}}_k=[{\bf{f}}_{k,1},{\bf{f}}_{k,2},\cdots,{\bf{f}}_{k,D_k+1}]^{\rm T},
\end{equation}
where ${\bf{f}}_{k,1}={\bf{g}}_k$, ${\bf{f}}_{k,n+1}={\bf{g}}_{\Lambda_k(n)},n=1,2,\cdots,D_k$, and ${\bf{g}}_k = \left[g_{k,1},g_{k,2},\cdots,g_{k,M}\right]^{\mathrm T}$.
Then, the RZF precoding vector of user $k$ is given by \cite{Nguyen2008, Zhu2016}
\begin{equation}\label{eq7.1}
{\bf{w}}_k^{\mathrm{RZF}}=\frac{({\bf{F}}_k^{\rm H}{\bf{F}}_k+\alpha_k{\bf{I}})^{-1}{\bf{f}}_{k,1}}{\parallel ({\bf{F}}_k^{\rm H}{\bf{F}}_k+\alpha_k{\bf{I}})^{-1}{\bf{f}}_{k,1}\parallel},
\end{equation}
where $\alpha_k$ is a regularization parameter.

\begin{prop} \label{prop3}
\emph{
With RZF precoding in \eqref{eq7.1} and perfect CSI, the ergodic achievable rate of user $k$ is given by
\begin{equation}\label{eq7.3}
R_{k}^{\mathrm{RZF}} = \log_2\left( 1 + \frac{ E^{\mathrm{s}} }{\sum\limits_{l\in U_k}  E^{\mathrm{i}}(l) +\sigma^2}\right),
\end{equation}
where $E^{\mathrm{s}}\rightarrow \frac{-M \beta_k \mathcal{G}^2(\rho_k,\xi_k) E_k}{\frac{d}{d\xi_k}\mathcal{G}(\rho_k,\xi_k)}$ is the received signal power of user $k$, $E^{\mathrm{i}}(l) \rightarrow \frac{\beta_k E_l}
{{(1+\mathcal{G}(\rho_l,\xi_l))^2}}$ is the interference power received at user $k$ and caused by user $l$.
Moreover, $\rho_k=D_k/M$, $\xi_k=\alpha_k/M$, and $\mathcal{G}(\rho_k,\xi_k)$ can be evaluated in closed form \cite{Nguyen2008}
\begin{align}\label{eq7.4}
\!\mathcal{G}(\rho_k,\xi_k)\!=\!\frac{1}{2}\left[ \sqrt{\frac{(1\!-\!\rho_k)^2}{\xi_k^2}\!+\!\frac{2(1\!+\!\rho_k)}{\xi_k}\!+\!1}
\!+\! \frac{1\!-\!\rho_k}{\xi_k} \!-1 \right]\!.\!
\end{align}
}
\end{prop}
\begin{IEEEproof}
Please refer to Appendix~\ref{proof3}.
\end{IEEEproof}

If uniform transmit power allocation is adopted, the ergodic achievable rate of user $k$ in \eqref{eq7.3} reduces to
\begin{equation}\label{eq7.5}
R_{k}^{\mathrm{RZF,uni}} = \log_2\left( 1 + \frac{E^{\mathrm{s,uni}} }{\sum\limits_{l\in U_k}  E^{\mathrm{i,uni}}(l) + \sigma^2}\!\right),
\end{equation}
where $E^{\mathrm{s,uni}}\rightarrow \frac{-M \beta_k \mathcal{G}^2(\rho_k,\xi_k) E_0}{\frac{d}{d\xi_k}\mathcal{G}(\rho_k,\xi_k){\overline{K}}}$, $E^{\mathrm{i,uni}}(l) \rightarrow \frac{\beta_k E_0}
{{(1+\mathcal{G}(\rho_l,\xi_l))^2}{\overline{K}}}$.
Similarly, the ergodic achievable rate of user $k$ in a \emph{conventional} massive MIMO system (i.e., $\rho_k=(K-1)/M$, $\xi_k=\alpha/M$, $\overline{K}=K$) can be obtained from \eqref{eq7.5}, and is given by
\begin{equation}\label{eq7.6}
R_{k}^{\mathrm{b,RZF,uni}}=\log_2\left(1+\frac{ E^{\mathrm{b,s,uni}}}{\sum\limits_{l\neq k} E^{\mathrm{b,i,uni}}(l) + \sigma^2}\right),
\end{equation}
where $E^{\mathrm{b,s,uni}}\rightarrow \frac{-M \beta_k \mathcal{G}^2(\frac{K-1}{M},\frac{\alpha}{M}) E_0}{\frac{d}{d\frac{\alpha}{M}}\mathcal{G}(\frac{K-1}{M},\frac{\alpha}{M}){{K}}}$,
$E^{\mathrm{b,i,uni}}(l) \rightarrow \frac{ \beta_k E_0}
{{(1+\mathcal{G}(\frac{K-1}{M},\frac{\alpha}{M}))^2}{{K}}}$, and $\alpha$ is the regularization parameter in \emph{conventional} massive MIMO.

\emph{Remark} 3.
Comparing $R_{k}^{\mathrm{RZF,uni}}$ with $R_{k}^{\mathrm{b,RZF,uni}}$, we observe that, by employing the proposed scheme, caching not only improves the transmit power by reducing the number of active users $\overline{K}$, but also impacts the signal and interference powers.
Therefore, for RZF precoding, the tradeoff between the signal and the interference powers, which is adjusted by the regularization parameter, has to be newly investigated for maximization of the cache-enabled performance gains. In this paper, the optimal regularization parameter is found numerically for the results shown in Section~\ref{Simulation Results}.

\section{Performance Evaluation} \label{Simulation Results}
In this section, we evaluate the performance of the proposed scheme. For comparison, \emph{conventional} massive MIMO is adopted as a baseline. Let $\mathrm{SNR} \triangleq 10\log_{10}\frac{E_0}{\sigma^2}$ be the transmit signal-to-noise ratio. We set $\mathrm{SNR} = 10$ dB, $\beta_k = 0.5$, $\forall k$, $F=1$ MByte, and $L_b=100$.
%Note that, owing to the law of large numbers, the actual realization of the random variables $\overline{K}$, $N_k$ (i.e., the number of interfering users for user $k$), and $D_k$ (i.e., the number of active users that do not have user $k$'s requested file in their caches) converge to their mean values  $\mathcal{E}\{\overline{K}\}$, $\mathcal{E}\{N_k\}$, and $\mathcal{E}\{D_k\}$, respectively, for large $K$. Thus, we set $\overline{K}=\mathcal{E}\{\overline{K}\}$, $N_k=\mathcal{E}\{N_k\}$, and $D_k=\mathcal{E}\{D_k\}$, respectively.
We assume that the files are requested with equal probability ${1}/{L_b}$. To illustrate the benefits of caching, we consider a simple uniform caching scheme, whereby each user caches a file with probability $p \triangleq {L_u}/{L_b}$.
We note that, by considering random requests and caching, $\overline{K}$, $N_k$ (i.e., the number of interfering users for user $k$), and $D_k$ (i.e., the number of active users that do not have user $k$'s requested file in their caches) become random variables.
However, for performance evaluation, we consider the case of asymptotically large $K$. For large $K$, due to the law of large numbers, $\overline{K}$, $N_k$, and $D_k$ converge to their mean values $\mathcal{E}\{\overline{K}\}$, $\mathcal{E}\{N_k\}$, and $\mathcal{E}\{D_k\}$, respectively.
Hence, we have $\mathcal{E}\{\overline{K}\}=(1-p)K$.
Furthermore, for a given user $k$, the event that user $l \neq k$ causes interference to user $k$, i.e., $l\in U_k$, has probability
\begin{equation}\label{eq9.1}
p_u=\Pr(c_{l,l}=1, c_{k,k}=1, c_{l,k}=1, l\neq k)=p_u^{(1)}+p_u^{(2)},
\end{equation}
where $p_u^{(1)}$ is the probability that users $l$ and $k$ request different files, which is given by
$p_u^{(1)}=C_{L_b}^1C_{L_b-1}^1 \big[ \frac{1}{L_b}(1-p) \big]^2(1-p)$. $p_u^{(2)}$ is the probability that the two users require the same file, given by $p_u^{(2)}=C_{L_b}^1 \big[ \frac{1}{L_b}(1-p) \big]^2$.
Consequently, $p_u=\big(1-\frac{1}{L_b}\big)(1-p)^3 + \frac{1}{L_b}(1-p)^2$.
Since $p_u$ is independent of the users, we have $\mathcal{E}\{N_k\}=(K-1)p_u$. Following the same approach as for $\mathcal{E}\{N_k\}$, we can further show $\mathcal{E}\{D_k\}=(K-1)p_u$. To show the maximum performance of RZF precoding, the regularization parameter is optimized numerically for each parameter setting.

\begin{figure}[!tbp]
\centering
\includegraphics[scale=0.5]{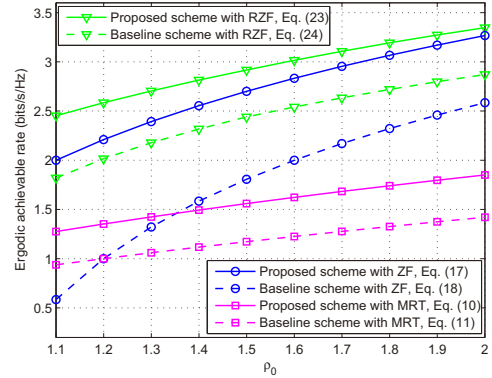}
\vspace{-0.5em}
\caption{Ergodic achievable rate per user vs. number of BS antennas per user, $\rho_0$, for MRT, ZF, and RZF precoders with $L_u=20$, i.e., $20\%$ of the users' requests are offloaded by caching.}
\label{fig1.3}
\vspace{-.2cm}
\end{figure}

\begin{figure}[!tbp]
\centering
\includegraphics[scale=0.5]{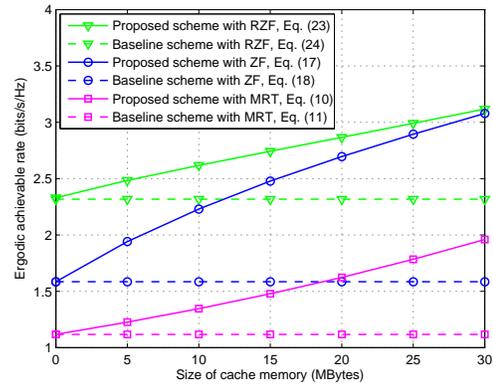}
\vspace{-0.5em}
\caption{Ergodic achievable rate per user vs. cache size, $L_uF$, for MRT, ZF, and RZF precoders with $\rho_0=1.4$.}
\label{fig1.5}
\vspace{-.4cm}
\end{figure}

Fig.~\ref{fig1.3} illustrates the ergodic achievable rate\footnote{The numerical results shown in this section were obtained with \eqref{eq2.8}, \eqref{eq2.6.3}, \eqref{eq3.9}, \eqref{eq3.3.1}, \eqref{eq7.5}, and \eqref{eq7.6} for $M, K \rightarrow \infty$, $\rho_0=M/K$, and have been validated by Monte Carlo simulations. However, for clarity, the simulation results are not included in Figs.~\ref{fig1.3} and~\ref{fig1.5}.} per user versus the number of BS antennas per user, $\rho_0$, for MRT, ZF precoding, and RZF precoding. From Fig.~\ref{fig1.3} we observe that, for all considered precoders, the proposed scheme achieves significantly higher ergodic rates than the baseline scheme. This is because, on the one hand, caching offloads the cellular traffic for inactive users and mitigates the multiuser interference of active users. On the other hand, with the enhanced precoders, caching is further exploited to improve the received signal power and/or increase the spatial degrees of freedom, both of which increase the ergodic achievable rate. For example, with MRT precoding, the proposed scheme for $\rho_0 =  1.1$ achieves the same performance as the baseline scheme for $\rho_0 =1.8$. Moreover, by optimally balancing between interference cancellation and power enhancement, enabled by user-side caching and BS-side precoding, respectively, the proposed scheme with RZF precoding achieves the best performance.

From Fig.~\ref{fig1.3} we also observe that the proposed scheme achieves the largest performance gains over the baseline scheme for ZF precoding when the number of antennas approaches the number of users. This is because the cache-enabled interference cancellation is exploited at the BS to reduce the number of ZF precoding constraints. For a small $\rho_0$, the signal space for ZF precoding design is severely constrained. In this case, ZF precoding can benefit from the increased spatial degrees of freedom enabled by caching and hence achieve a large performance improvement. Thus, caching can effectively enhance the performance of massive MIMO systems having a small number of BS antennas per user, i.e., when $\rho_0$ is small.

Fig.~\ref{fig1.5} shows the ergodic achievable rate per user versus the cache size, $L_uF$, for MRT, ZF, and RZF precoding. We observe that, for all considered precoders, as the cache size increases, the proposed scheme can exploit the increased offloading and interference cancellation opportunities enabled by caching to significantly improve the system performance.
For example, when $L_u = 20$, we have $\mathcal{E}\{\overline{K}\}/K = 0.8$, i.e., $20\%$ of the users' requests are offloaded by caching, and $p_u=\mathcal{E}\{N_k\}/(K-1) =\mathcal{E}\{D_k\}/(K-1) = 0.513$, i.e., $28.7\%$ of the users have cached the files requested by other users. Consequently, due to caching, $48.7\%$ of the users including the offloaded users would not cause interference to the active users.
In this case, for ZF precoding, the ergodic achievable rate of the proposed scheme increases by $70.1\%$  compared to the baseline scheme.
%say how much traffic can be offloaded and how much interference can be cancelled as we did in the previous version. Also say how much performance gains can be achieved, e.g., by ZF, in this example.
On the other hand, as caching is unavailable for the baseline scheme, its performance is independent of $L_u$.
From Fig.~\ref{fig1.5} we also observe that, for the proposed scheme, RZF precoding achieves the best performance among the considered precoding techniques for all considered cache sizes.

%Interestingly, the performance gap between the proposed and the baseline schemes with ZF precoding is much larger than that with MRT, for all considered $\gamma$. This is because, with ZF precoding, cache-enabled offloading not only enhances the power

%On the other hand, For RZF precoding, a stable high performance for all considered $\gamma$

%\newpage

% there exists an optimal $\gamma$ for maximization of the ergodic achievable rate. This is because, when $\gamma$ is small, although RZF precoding introduces only a small amount of interference, the system suffers from a loss of received power. In contrast, when $\gamma$ is large, RZF enhances the received power but introduces a large interference. On the other hand, using the optimal $\gamma$, RZF balances between interference cancellation and power enhancement and hence, achieves the optimal performance. Moreover, as $\rho_0$ increases, the impact of interference reduces and hence, the optimal values of $\gamma$ for both schemes enlarge. From Fig.~\ref{fig1.5} we also observe that the proposed scheme achieves significantly larger ergodic rate than the baseline scheme, as caching enables additional power enhancement and interference cancellation.

\section{Conclusion} \label{Conclusion}
In this paper, a novel cache-aided massive MIMO scheme was proposed. In addition to reaping the advantages of caching and massive MIMO, the proposed scheme also facilitates interference cancellation at the user side and transmit power savings at the BS. Exploiting these cache-enabled benefits, linear precoders, specifically MRT, ZF, and RZF precoders, were redesigned for further performance improvement. Closed-form expressions for the ergodic achievable rate of the proposed schemes were derived for MRT, ZF, and RZF precoding. Numerical results show that the proposed scheme significantly improves the performance of all considered precoding techniques especially when the number of BS antennas per user is small.

\section*{Acknowledgment}
This work was supported in part by National Science Foundation of China with Grant numbers 61771216 and 61531011. The work of L. Xiang is supported by the project FNR CORE ProCAST, grant R-AGR-3415-10.

\begin{appendix}
\subsection{Proof of Proposition~\ref{prop1}} \label{proof1}
Substituting \eqref{eq2.6} into \eqref{eq2.5}, we have
\begin{equation}\label{eqp1.1}
\tilde{R}_{k}=\log_2\left(1+\left(\mathcal{E}\left\{\frac{\sum\limits_{l\in U_k}
\frac{\left| {\bf{h}}_{k}^{\mathrm{H}}{\mathbf{h}}_{l} \right|^2}{ \left\| {\mathbf{h}}_l \right\|^2} E_l+\sigma^2}
{ \left\| {\mathbf{h}}_{k} \right\|^2 E_k}\right\}\right)^{-1}\right),
\end{equation}
where \cite[Appendix A]{N2013},
\begin{align}\label{eqp1.3}
\!\!\mathcal{E}\!\left\{\!\frac{\sum\limits_{l\in U_k}\!\!
\frac{ \left| {\mathbf{h}}_{k}^{\mathrm{H}}{\mathbf{h}}_{l} \right|^2}{ \left\| {\mathbf{h}}_l \right\|^2} E_l\!+\!\sigma^2}
{ \left\| {\mathbf{h}}_{k} \right\|^2 E_k}\!\right\}
\!\!=\!\! \left(\! \sum\limits_{l \in U_k\!\!\!\!\!} \beta_k E_l \!+\! \sigma^2\!\right) \!\mathcal{E}\!\left\{\!\frac{1}
{ \left\| {\mathbf{h}}_{k} \right\|^2 \!E_k}\!\right\}\!.\!\!
\end{align}
Due to \eqref{eq1.2}, we have $ \left\|{\mathbf{h}}_{k} \right\|^2=\beta_k\sum\nolimits_{m=1}^M  \left| g_{k,m}  \right|^2$. Note that, $ \left| g_{k,m} \right| ^2 = \left| \Re \left(g_{k,m} \right) \right|^2+ \left| \Im \left(g_{k,m} \right) \right|^2$.
Hence, $2\sum\nolimits_{m=1}^M  \left| g_{k,m} \right| ^2$ follows the chi-squared distribution with $2M$ degrees of freedom as $\sqrt{2} \Re \left(g_{k,m}\right)$ and $\sqrt{2} \Im \left(g_{k,m} \right)$ are independent standard normal random variables.
Thus, $2\sum\nolimits_{m=1}^M  \left|g_{k,m} \right| ^2$ is an inverse chi-square distribution with $2M$ degrees of freedom, and we have $\mathcal{E}\Big\{\left(\sum\nolimits_{m=1}^M  \left| g_{k,m} \right|^2\right)^{-1}\Big\} = (M-1)^{-1}$, and $\mathcal{E}\left\{{1} / { \left\| {\mathbf{h}}_{k} \right\| ^2}\right\}=\frac{1}{\beta_k(M-1)}$.
Substituting \eqref{eqp1.3} into \eqref{eqp1.1}, Proposition~\ref{prop1} is proved.

\subsection{Proof of Proposition~\ref{prop2}} \label{proof2}
Substituting \eqref{eq3.5} into \eqref{eq2.4}, we have
\begin{equation}\label{eqp2.1}
R_{k}^{\mathrm{ZF}}=\mathcal{E}\left\{\log_2\left(1+ \frac{E_k}{\sigma^2} \frac{ \left| {\mathbf{h}}_{k}^{\mathrm{H}} {\mathbf{Q}}_k({\mathbf{Q}}_k^{\mathrm{H}}{\mathbf{Q}}_k)^{-1}{\mathbf{e}}_1  \right|^2 }{ \left\| {\mathbf{Q}}_k({\mathbf{Q}}_k^{\mathrm{H}}{\mathbf{Q}}_k)^{-1}{\mathbf{e}}_1 \right\| ^2} \right)\right\}.
\end{equation}
Since ${ |{\mathbf{q}}_{1}^{\mathrm{H}}{\mathbf{Q}}_k \left({\mathbf{Q}}_k^{\mathrm{H}}{\mathbf{Q}}_k \right)^{-1}{\mathbf{e}}_l |^2} =0$, $l = 2,\cdots,D_k+1$, we have ${ |{\mathbf{q}}_{1}^{\mathrm{H}}{\mathbf{Q}}_k \left({\mathbf{Q}}_k^{\mathrm{H}}{\mathbf{Q}}_k \right)^{-1}{\mathbf{e}}_1  |^2} = { \|{\mathbf{q}}_{1}^{\mathrm{H}}{\mathbf{Q}}_k \left({\mathbf{Q}}_k^{\mathrm{H}}{\mathbf{Q}}_k \right)^{-1} \|^2} = { {\mathbf{q}}_{1}^{\mathrm{H}}{\mathbf{Q}}_k \left({\mathbf{Q}}_k^{\mathrm{H}}{\mathbf{Q}}_k \right)^{-1} \left({\mathbf{Q}}_k^{\mathrm{ H}}{\mathbf{Q}}_k\right)^{-1}{\mathbf{Q}}_k^{\mathrm{H}}{\mathbf{q}}_1 }$. Note that ${{\mathbf{q}}_{1}^{\mathrm{H}}{\mathbf{Q}}_k({\mathbf{Q}}_k^{\mathrm{H}}{\mathbf{Q}}_k)^{-1}({\mathbf{Q}}_k^{\mathrm{H}}{\mathbf{Q}}_k)^{-1}{\mathbf{Q}}_k^{\mathrm{H}}{\mathbf{q}}_1 }$ is the element in the first row and the first column of
${{\mathbf{Q}}_k^{\mathrm{H}}{\mathbf{Q}}_k ({\mathbf{Q}}_k^{\mathrm{H}}{\mathbf{Q}}_k )^{-1} ({\mathbf{Q}}_k^{\mathrm{H}}{\mathbf{Q}}_k )^{-1}{\mathbf{Q}}_k^{\mathrm{H}}{\mathbf{Q}}_k} = {\mathbf{I}}$. Hence, we have ${ |{\mathbf{h}}_{k}^{\mathrm{H}}{\mathbf{Q}}_k\left({\mathbf{Q}}_k^{\mathrm{H}}{\mathbf{Q}}_k\right)^{-1}{\bf{e}}_1 |^2} = 1$ as ${\mathbf{h}}_k = {\mathbf{q}}_1$. Then, substituting ${ |{\mathbf{h}}_{k}^{\mathrm{H}}{\mathbf{Q}}_k \left({\mathbf{Q}}_k^{\mathrm{H}}{\mathbf{Q}}_k \right)^{-1}{\mathbf{e}}_1 |^2} = 1$ into \eqref{eqp2.1}, \eqref{eq3.3} in Proposition~\ref{prop2} is proved.

Moreover, substituting \eqref{eq3.3} into \eqref{eq2.5}, the downlink achievable rate of user $k$ is lower bounded by
\begin{align}\label{eqp3.1}
\tilde{R}_{k}^{\mathrm{ZF}}
\!=\!\log_2\left(1+ \frac{E_k}{\sigma^2} \left(\mathcal{E}\left\{ {{ \left\|{\mathbf{Q}}_k({\mathbf{Q}}_k^{\mathrm{H}}{\mathbf{Q}}_k)^{-1}{\mathbf{e}}_1  \right\|^2} }  \right\}\right)^{-1}\right).
\end{align}
%We have
%\begin{align}\label{eqp3.2}
%\mathcal{E}\left\{ \frac{{\parallel \! {\bf{Q}}_k({\bf{Q}}_k^{\rm H}{\bf{Q}}_k)^{-1}{\bf{e}}_1 \!\parallel^2}\sigma^2}{E_k} \right\}
%\!=\!\frac{\sigma^2}{E_k}
%\mathcal{E}\left\{ {\parallel \! {\bf{Q}}_k({\bf{Q}}_k^{\rm H}{\bf{Q}}_k)^{-1}{\bf{e}}_1 \!\parallel^2} \right\}.
%\end{align}
Since ${ \| {\bf{Q}}_k({\bf{Q}}_k^{\rm H}{\bf{Q}}_k)^{-1}{\bf{e}}_1 \|^2}=  {\bf{e}}_1^{\rm H}({\bf{Q}}_k^{\rm H}{\bf{Q}}_k)^{-1}{\bf{Q}}_k^{\rm H}{\bf{Q}}_k({\bf{Q}}_k^{\rm H}{\bf{Q}}_k)^{-1}{\bf{e}}_1  =  {\bf{e}}_1^{\rm H}({\bf{Q}}_k^{\rm H}{\bf{Q}}_k)^{-1}{\bf{e}}_1 $, we have that ${ \| {\bf{Q}}_k({\bf{Q}}_k^{\rm H}{\bf{Q}}_k)^{-1}{\bf{e}}_1 \|^2}$ is the element in the first row and the first column of matrix $({{\bf{Q}}_k^{\rm H}{\bf{Q}}_k})^{-1}$. Note that $({{\bf{Q}}_k^{\rm H}{\bf{Q}}_k})^{-1}$ is a complex inverse Wishart matrix with $D_k+1$ degrees of freedom and parameter matrix ${\bf{\Phi}}^{-1}$ \cite{Tulino2004}, where ${\bf{\Phi}}\!\in\!{{\mathbb C}^{{(D_k+1)} \times {(D_k+1)}}}$ is a diagonal matrix with diagonal elements $[\beta_k, \beta_{\Lambda_{k} (1)}, \cdots, \beta_{\Lambda_k(D_k)}]$. %$[\rho_1,\rho_2,\cdots,\rho_{D_k+1}]$.
Using the results in \cite[Ch. 3.8]{Mardia1979}, we have $\mathcal{E}\left\{ ({{\bf{Q}}_k^{\rm H}{\bf{Q}}_k})^{-1} \right\}=\frac{{\bf{\Phi}}^{-1}}{M-(D_k+1)}$. Thus, we have
\begin{align}\label{eqp3.3}
\mathcal{E}\left\{ { \left\| {\bf{Q}}_k({\bf{Q}}_k^{\rm H}{\bf{Q}}_k)^{-1}{\bf{e}}_1 \right\| ^2} \right\}
%&=\frac{1}{\left(M-D_k-1\right)\rho_1} \nonumber \\
&=\frac{1}{\left(M-D_k-1\right)\beta_k}.
\end{align}
Then, substituting \eqref{eqp3.3} into \eqref{eqp3.1}, \eqref{eq3.8} in Proposition~\ref{prop2} is proved.

\subsection{Proof of Proposition~\ref{prop3}} \label{proof3}
Based on \eqref{eq2.3} and \eqref{eq7.1}, the effective signal power is given as
\begin{align}\label{eq6.2}
E^{\mathrm{s}}&= \left|{\bf{h}}_k^{\rm H}{\bf{w}}_k^{\mathrm{RZF}} \right|^2 E_k
=\frac{ \left| {\bf{h}}_k^{\rm H} ({\bf{F}}_k^{\rm H}{\bf{F}}_k+\alpha_k{\bf{I}})^{-1}{\bf{f}}_{k1} \right|^2E_k}
{ \left\| ({\bf{F}}_k^{\rm H}{\bf{F}}_k+\alpha_k{\bf{I}})^{-1}{\bf{f}}_{k1} \right\|^2} \nonumber \\
&=\frac{ \left| {\bf{g}}_k^{\rm H} ({\bf{F}}_k^{\rm H}{\bf{F}}_k+\alpha_k{\bf{I}})^{-1}{\bf{g}}_{k} \right|^2 \beta_k E_k}
{ \left\| ({\bf{F}}_k^{\rm H}{\bf{F}}_k+\alpha_k{\bf{I}})^{-1}{\bf{g}}_{k} \right\|^2}.
\end{align}
Applying the matrix inversion lemma \cite{Nguyen2008}, we have
\begin{align}\label{eq6.3}
({\bf{F}}_k^{\rm H}{\bf{F}}_k+\alpha_k{\bf{I}})^{-1}{\bf{g}}_{k}
&\!=\!\frac{({\bf{F}}_{k(k)}^{\rm H}{\bf{F}}_{k(k)}+\alpha_k{\bf{I}})^{-1}{\bf{g}}_{k}}
{1+{\bf{g}}_k^{\rm H} ({\bf{F}}_{k(k)}^{\rm H}{\bf{F}}_{k(k)}+\alpha_k{\bf{I}})^{-1}{\bf{g}}_{k} },
\end{align}
where ${\bf{F}}_{k(k)}$ is obtained by deleting the vector ${\bf{g}}_k$ from ${\bf{F}}_{k}$.
Defining
\begin{align}\label{eq6.4}
X_{k}={\bf{g}}_k^{\rm H} ({\bf{F}}_{k(k)}^{\rm H}{\bf{F}}_{k(k)}+\alpha_k{\bf{I}})^{-1}{\bf{g}}_{k},
\end{align}
\begin{align}\label{eq6.5}
{\bf{\Phi}}_{k}=({\bf{F}}_{k(k)}^{\rm H}{\bf{F}}_{k(k)}+\alpha_k{\bf{I}})^{-1},\quad \quad \:
\end{align}
with \eqref{eq6.3}, we have
\begin{align}\label{eq6.6}
E^{\mathrm{s}}&=\frac{\mid X_{k} \mid^2 \beta_k E_k}
{{ \left\| {\bf{\Phi}}_{k}{\bf{g}}_k \right\|^2}}.
\end{align}
%where the second equality holds for $M, D_k \rightarrow \infty$.

Rewrite $X_{k}$ as $X_{k}=\frac{1}{M}{\bf{g}}_k^{\rm H} (\frac{1}{M}{\bf{F}}_{k(k)}^{\rm H}{\bf{F}}_{k(k)}+\xi_k{\bf{I}})^{-1}{\bf{g}}_{k}$, where $\xi_k={\alpha_k}/{M}$. Then, in the large system limit where $M, D_k \rightarrow \infty$, but $\rho_k={D_k}/{M}$ is finite and fixed,
$X_k$ converges (almost surely) to \cite{Nguyen2008, Evans2000}
\begin{align}\label{eq6.7}
\mathcal{G}(\rho_k,\xi_k)=\int_0^{\infty}\frac{1}{\mu +\xi_k}d\mathcal{F}_{\rho_k}(\mu),
\end{align}
where
\begin{align}\label{eq6.8}
\mathcal{F}_{\rho_k}(\mu)&\stackrel{\Delta}{=}(1\!-\!\rho_k)^+\delta(\mu) \nonumber \\ &+\frac{\sqrt{(\mu-(1-\sqrt{\rho_k})^2)^+((1+\sqrt{\rho_k})^2-\mu)^+}}{2\pi \mu},
\end{align}
$\delta(\mu)$ is the impulse function, and $(x)^+ \triangleq \max \{0,x\}$.
%\begin{align}\label{eq6.7}
%\mathcal{G}(\rho_k,\xi_k)=\int_0^{\infty}\frac{1}{\mu +\xi_k}\mathcal{F}_{\rho_k}(\mu)d\mu,
%\end{align}
%where
%\begin{align}\label{eq6.8}
%\mathcal{F}_{\rho_k}(\mu)&=(1\!-\!\rho_k)^+\delta(\mu) \nonumber \\ &+\frac{\sqrt{(\mu-(1-\sqrt{\rho_k})^2)^+((1+\sqrt{\rho_k})^2-\mu)^+}}{2\pi \mu}.
%\end{align}
On the other hand, define $\bar{\bf{\Phi}}_k\!\!=\!\!(\frac{1}{M}{\bf{F}}_{k(k)}^{\rm H}{\bf{F}}_{k(k)}+ \xi_k {\bf{I}})^{-1}$.
Then, we have
\begin{align}\label{eq6.10}
{{ \left\| {\bf{\Phi}}_{k}{\bf{g}}_k \right\|^2}}
={{ \left| {\bf{g}}_k^{\rm H}{\bf{\Phi}}_{k}^{\rm H}{\bf{\Phi}}_{k}{\bf{g}}_k \right|}}
=\frac{1}{M}{{ | \frac{1}{M} {\bf{g}}_k^{\rm H}\bar{\bf{\Phi}}_k\bar{\bf{\Phi}}_k{\bf{g}}_k |}}.
\end{align}
Following a similar approach as for $X_k$, one can show that
\begin{align}
{\frac{1}{M} {\bf{g}}_k^{\rm H}\bar{\bf{\Phi}}_k\bar{\bf{\Phi}}_k{\bf{g}}_k }
& \rightarrow
\int_0^{\infty}\frac{1}{(\mu +\xi_k)^2}\mathcal{F}_{\rho_k}(\mu)d\mu \label{eq6.10b} \\
&\rightarrow
-\frac{d}{d\xi_k}\mathcal{G}(\rho_k,\xi_k), \label{eq6.11}
\end{align}
where the second line holds since $\frac{1}{(\mu +\xi_k)^2}=-\frac{d}{d\xi_k}\frac{1}{(\mu +\xi_k)}$. %, we have
%\begin{align}\label{eq6.11}
%{\frac{1}{M} {\bf{g}}_k^{\rm H}\bar{\bf{\Phi}}_k\bar{\bf{\Phi}}_k{\bf{g}}_k }
%\rightarrow
%-\frac{d}{d\xi_k}\mathcal{G}(\rho_k,\xi_k).
%\end{align}
Substituting \eqref{eq6.7} and \eqref{eq6.11} into \eqref{eq6.6}, we get
\begin{align}\label{eq6.12}
E^{\mathrm{s}}\rightarrow
\frac{\mathcal{G}^2(\rho_k,\xi_k)\beta_kE_k}
{-\frac{1}{M}\frac{d}{d\xi_k}\mathcal{G}(\rho_k,\xi_k)}
=\frac{-M\mathcal{G}^2(\rho_k,\xi_k)\beta_kE_k}
{\frac{d}{d\xi_k}\mathcal{G}(\rho_k,\xi_k)}.
\end{align}

If $l\in U_k$, the effective interference power of user $l$ to user $k$ is given by
\begin{align}\label{eq6.13}
E^{\mathrm{i}}(l)&=| {\bf{h}}_k^{\rm H}{\bf{w}}_l^{\mathrm{RZF}}|^2 E_l
=\frac{| {\bf{h}}_k^{\rm H} ({\bf{F}}_l^{\rm H}{\bf{F}}_l+\alpha_l{\bf{I}})^{-1}{\bf{f}}_{l,1} |^2E_l}
{\| ({\bf{F}}_l^{\rm H}{\bf{F}}_l+\alpha_l{\bf{I}})^{-1}{\bf{f}}_{l,1}\|^2} \nonumber \\
&=\frac{| {\bf{g}}_k^{\rm H} ({\bf{F}}_l^{\rm H}{\bf{F}}_l+\alpha_l{\bf{I}})^{-1}{\bf{g}}_{l} |^2 \beta_k E_l}
{\| ({\bf{F}}_l^{\rm H}{\bf{F}}_l+\alpha_l{\bf{I}})^{-1}{\bf{g}}_{l}\|^2}.
\end{align}
Applying the matrix inversion lemma, \eqref{eq6.13} is rewritten as
\begin{align}\label{eq6.14}
E^{\mathrm{i}}(l)=\frac{| {\bf{g}}_k^{\rm H} {\bf{\Phi}}_l {\bf{g}}_l |^2 \beta_k E_l}
{{\| {\bf{\Phi}}_{l}{\bf{g}}_l\|^2}}.
\end{align}
Removing ${\bf{g}}_k$ from ${\bf{F}}_{l(l)}$ and applying the matrix inversion lemma, we have
\begin{align}\label{eq6.17}
| {\bf{g}}_k^{\rm H} {\bf{\Phi}}_l {\bf{g}}_{l} |^2
=\frac{| {\bf{g}}_k^{\rm H}({\bf{F}}_{l(lk)}^{\rm H}{\bf{F}}_{l(lk)}+\alpha_l{\bf{I}})^{-1}{\bf{g}}_{l}|^2}
{| 1+{\bf{g}}_k^{\rm H} ({\bf{F}}_{l(lk)}^{\rm H}{\bf{F}}_{l(lk)}+\alpha_l{\bf{I}})^{-1}{\bf{g}}_{k} |^2},
\end{align}
where ${\bf{F}}_{l(lk)}$ is obtained by deleting vectors ${\bf{g}}_k$ and ${\bf{g}}_l$ from ${\bf{F}}_{l}$.
Therein, for the numerator, we have
\begin{align}\label{eq6.17a}
&{| {\bf{g}}_k^{\rm H}{\bf{\Phi}}_{l(k)}{\bf{g}}_{l}|^2}
=\frac{1}{M^2}{\bf{g}}_k^{\rm H} {\bar{\bf{\Phi}}}_{l(k)} {\bf{g}}_l {\bf{g}}_l^{\rm H} {\bar{\bf{\Phi}}}_{l(k)}^{\rm H} {\bf{g}}_k \nonumber \\
&\rightarrow \frac{1}{M}\mathrm{tr}( \frac{1}{M} {\bar{\bf{\Phi}}}_{l(k)} {\bf{g}}_l {\bf{g}}_l^{\rm H}  {\bar{\bf{\Phi}}}_{l(k)}^{\rm H} )  =\frac{1}{M^2} {\bf{g}}_l^{\rm H} {\bar{\bf{\Phi}}}_{l(k)}^{\rm H}{\bar{\bf{\Phi}}}_{l(k)} {\bf{g}}_l  \nonumber \\
&\rightarrow \frac{1}{M} \int_0^{\infty}\frac{1}{(\mu +\xi_l)^2}\mathcal{F}_{\rho_l^-}(\mu)d\mu
\rightarrow -\frac{1}{M}\frac{d}{d\xi_l}\mathcal{G}(\rho_l^-,\xi_l).
\end{align}
where $\rho_l^-={(N_l-1)}/{M}$,
${\bf{\Phi}}_{l(k)}=({\bf{F}}_{l(lk)}^{\rm H}{\bf{F}}_{l(lk)}+\alpha_l{\bf{I}})^{-1}$, and ${\bar{\bf{\Phi}}}_{l(k)}=(\frac{1}{M}{\bf{F}}_{l(lk)}^{\rm H}{\bf{F}}_{l(lk)}+\frac{\alpha_l}{M}{\bf{I}})^{-1}$.
Moreover, for the denominator, based on (\ref{eq6.7}), we have
\begin{align}\label{eq6.17b}
{| 1 \!+\!{\bf{g}}_k^{\rm H} ({\bf{F}}_{l(lk)}^{\rm H}{\bf{F}}_{l(lk)}+\alpha_l{\bf{I}})^{-1}{\bf{g}}_{k} |^2}
\rightarrow
{(1+\mathcal{G}(\rho_l^-,\xi_l))^2}.
\end{align}
As $M, N_l \rightarrow \infty$, we have $\rho_l^-\rightarrow \rho_l$, and hence,
substituting (\ref{eq6.17a}) and (\ref{eq6.17b}) into (\ref{eq6.17}), we get that
\begin{align}\label{eq6.20}
| {\bf{g}}_k^{\rm H} {\bf{\Phi}}_l {\bf{g}}_{l} |^2
\rightarrow
\frac{-\frac{1}{M}\frac{d}{d\xi_l}\mathcal{G}(\rho_l,\xi_l)}
{(1+\mathcal{G}(\rho_l,\xi_l))^2}.
\end{align}
Based on (\ref{eq6.11}), we have
%\begin{align}\label{eq6.15}
%{{\| {\bf{\Phi}}_{l}{\bf{g}}_l\|^2}}
%&={\bf{g}}_l^{\rm H} {\bf{\Phi}}_{l}^{\rm H} {\bf{\Phi}}_{l} {\bf{g}}_l
%=\frac{1}{M}\left(\frac{1}{M} {\bf{g}}_l^{\rm H} {\bar{\bf{\Phi}}}_{l}^{\rm H} {\bar{\bf{\Phi}}}_{l} {\bf{g}}_l\right) \nonumber \\
%&\rightarrow -\frac{1}{M}\frac{d}{d\xi_l}\mathcal{G}(\rho_l,\xi_l),
%\end{align}
\begin{align}\label{eq6.15}
{{\| {\bf{\Phi}}_{l}{\bf{g}}_l\|^2}}
=\frac{1}{M}\left(\frac{1}{M} {\bf{g}}_l^{\rm H} {\bar{\bf{\Phi}}}_{l}^{\rm H} {\bar{\bf{\Phi}}}_{l} {\bf{g}}_l\right)
\rightarrow -\frac{1}{M}\frac{d}{d\xi_l}\mathcal{G}(\rho_l,\xi_l),
\end{align}
which holds for $M \rightarrow \infty$.
Hence, we have
\begin{align}\label{eq6.18}
E^{\mathrm{i}}(l) \rightarrow \frac{\beta_k E_l}
{{(1+\mathcal{G}(\rho_l,\xi_l))^2}}.
\end{align}

Moreover, the noise power is $\sigma^2$.
Hence, when $D_k, M \rightarrow \infty$, but $\rho_k$ is finite and fixed, the $\mathrm{SINR}$ of user $k$ is given by
\begin{align}\label{eq6.19}
\mathrm{SINR}_k=\frac{ E^{\mathrm{s}}}{\sum\nolimits_{l \in U_k} E^{\mathrm{i}}(l)+\sigma^2}.
\end{align}
Substituting (\ref{eq6.12}), (\ref{eq6.18}), and (\ref{eq6.19}) into (\ref{eq2.4}), Proposition~\ref{prop3} is proved.

\end{appendix}

\end{document}